\documentclass[twocolumn]{autart}                                                                                                                   \usepackage{graphicx}                                                                        
\usepackage{amsmath}
\usepackage{amsfonts}
\usepackage{times}

\usepackage{xcolor}

\usepackage{tikz}
\usetikzlibrary{shapes,arrows}
\tikzstyle{block} = [draw, fill=white, rectangle, minimum height=2em, minimum width=2.5em]
\tikzstyle{largeblock} = [draw, fill=white, rectangle, minimum height=3em, minimum width=3em]
\tikzstyle{smallblock} = [draw, fill=white, rectangle, minimum height=2em, minimum width=2.5em]
\tikzstyle{input} = [coordinate]
\tikzstyle{sum} = [draw, fill=white, circle, minimum size=3pt, inner sep=3pt, label={center:\tiny$+$}]
\tikzstyle{node} = [draw, fill=black, circle, minimum size=2pt, inner sep=0pt]
\tikzstyle{output} = [coordinate]
\tikzstyle{pinstyle} = [pin edge={to-,thin,black}]
\tikzstyle{line} = [draw, -latex']

\usepackage[normalem]{ulem}

\begin{document}

\begin{frontmatter}
                                                                                                                                          
\title{Passivity-Based Output-Feedback Control\\of Turbulent Channel Flow\thanksref{footnoteinfo}} 
\thanks[footnoteinfo]{This paper was presented in part at UKACC $10^{\mathrm{th}}$ International Conference on Control, Loughborough, U.K. July 2014. Corresponding author P.~H.~Heins. Tel. (+44) (0) 114 222 5613. }

\author[Sheffield]{Peter H. Heins}\ead{p.heins@sheffield.ac.uk},    \author[Sheffield]{Bryn Ll. Jones}\ead{b.l.jones@sheffield.ac.uk},               \author[Southampton]{Ati. S. Sharma}\ead{a.sharma@soton.ac.uk}  
\address[Sheffield]{Department of Automatic Control and Systems Engineering, University of Sheffield, Sheffield, S1 3JD, UK}  \address[Southampton]{Engineering and the Environment, University of Southampton, Highfield, Southampton, SO17 1BJ, UK}                    
          

\begin{abstract}                          This paper describes a robust linear time-invariant output-feedback control strategy to reduce turbulent fluctuations, and therefore skin-friction drag, in wall-bounded turbulent fluid flows, that nonetheless gives performance guarantees in the nonlinear turbulent regime. The novel strategy is effective in reducing the supply of available energy to feed the turbulent fluctuations, expressed as reducing a bound on the supply rate to a quadratic storage function. The nonlinearity present in the equations that govern the dynamics of the flow is known to be passive and can be considered as a feedback forcing to the linearised dynamics (a Lur'e decomposition). Therefore, one is only required to control the linear dynamics in order to make the system close to passive. The ten most energy-producing spatial modes of a turbulent channel flow were identified. Passivity-based controllers were then generated to control these modes. The controllers require measurements of streamwise and spanwise wall-shear stress, and they actuate via wall transpiration. Nonlinear direct numerical simulations demonstrated that these controllers were capable of significantly reducing the turbulent energy and skin-friction drag of the flow.

\end{abstract}

\end{frontmatter}

\section{Introduction}\label{intro}
Turbulent channel flows are characterised by their self-sustaining chaotic motions and are known to induce high skin-friction drag. Conversely, laminar channel flow has the lowest sustainable skin-friction drag~\cite{Bewley04} and is stable to infinitesimal perturbations (turbulent fluctuations) for $Re<5772$\footnote{$Re := \frac{\mathrm{U}_{cl} h }{\nu}$, centreline Reynolds number, where $\mathrm{U}_{cl}$ is the maximum laminar centreline velocity, $h$ is the channel half-height and $\nu$ is the kinematic viscosity.}\cite{Trefethen93}. However, experiments and simulations show that channel flow can sustain turbulence for $Re$ as low as 1000~\cite{SchmidHenningson} and transition to turbulence can occur at these low Reynolds numbers to perturbations of finite amplitude. This is because the nonlinearity in the Navier-Stokes equations plays a significant role, so a full consideration of stability must take it into account. When attempting to control turbulent wall-bounded flows, difficulties arise in several areas. The first of these is modelling. The equations that model the dynamics of all incompressible Newtonian fluids, the incompressible Navier-Stokes equations, are a set of nonlinear partial differential algebraic equations (PDAEs). The linear dynamics of turbulent fluids are known to be responsible for {all of the} energy production~\cite{schmid2007}. Therefore, it is justifiable to only control these dynamics as long as the effect of the nonlinearity is modelled appropriately as a source of uncertainty. In order to form a finite-dimensional linear control model, these equations must first be linearised around a known equilibrium solution and then discretised resulting in a set of differential algebraic equations. The scales of the flow that need to be controlled are not always known. Therefore, there {is} a balancing act of ensuring that the state dimension of the model is large enough to resolve the smaller scales but making it small enough so that controller synthesis is feasible. Adequate estimation and actuation are other issues. Practically, it is likely that sensors and actuators will be restricted to the walls. This limits the accuracy of flow estimations and efficacy of control actuation away from the walls. These issues combine to limit the drag reduction that can be achieved by output-feedback control.        \\
\\
There has been a significant amount of research into the use of modern control theory to reduce skin-friction drag in turbulent channel flow. A significant portion of this research has investigated the performance of state-feedback linear quadratic regulator~\cite{Bewley98,Hogberg03b,JLimThesis} and dynamic output-feedback linear quadratic Gaussian controllers~\cite{Lee01}. Model predictive control has also been used for drag reduction, both state-feedback~\cite{BewleyMoin01} and output-feedback controllers~\cite{Lee98} have been investigated. Furthermore, static output-feedback control laws have been derived capable of globally stabilising low-Reynolds number two-dimensional channel flows~\cite{Balogh01}. For an overview of the field of flow control in general, the reader is referred to the books by Aamo and Krsti\'c~\cite{Aamo}, Barbu~\cite{Barbu} and Gad-el-Hak~\cite{GadelHakBook}. \\
\\
There are many sources of uncertainty when controlling fluidic systems; examples include exogenous disturbances to flow variables, parametric uncertainty and modelling uncertainty. Passivity-based control has been proven to be both effective and robust to disturbance uncertainty. Sharma et al.~\cite{Sharma11} designed globally stabilising linear time-invariant (LTI) passivity-based controllers capable of relaminarising $Re_\tau$ = 100 channel flow\footnote{$Re_\tau = \frac{\mathrm{u}_\tau h}{\nu}$, skin-fiction velocity $\mathrm{u}_\tau = \sqrt{\frac{\tau_w}{\rho}}$, $\tau_w$ is wall-shear stress and $\rho$ is fluid density.}. Note that laminar incompressible channel flow has the lowest sustainable drag~\cite{Bewley04}. The controllers required full flow field information of the wall-normal velocity whilst actuation was via body-forcing on the wall-normal velocity throughout the channel. In their approach, they recognised that the nonlinearity in the Navier-Stokes equations acts a passive feedback operator. The passivity theorem states that two passive systems in feedback leads to the global system being passive. Therefore, it was required to only enforce passivity on the linear system to guarantee global stability. With the choice of sensing and actuation used, the controllers were capable of making the linear dynamics passive, but only just. It was found that only the four lowest spatial Fourier modes of the system needed to be controlled in order for flow relaminarisation to occur, suggesting that it is these modes that are most important for energy production. In an earlier work, Sharma~\cite{Sharma09} found that the passivity framework could also be applied to the linearised Navier-Stokes equations for the purposes of robust model reduction. \\
\\
The aim of this paper is to extend the work of Sharma et al.~\cite{Sharma11} towards passivity-based control of turbulent channel flow with actuation \emph{and} sensing restricted to the walls; in this particular case, sensing of streamwise and spanwise wall-shear stress and actuation via wall transpiration. This will show the drag reduction achievable by a passivity-based controller with more realistic sensing and actuation than that used previously. This is the first time an output-feedback passivity-based control method with wall sensing and actuation has been applied to turbulent flow. As will be demonstrated, with the sensing/actuation arrangement employed, it will not be possible to enforce passivity. Instead, passivity-based control shall be used to minimise the upper bound on energy production of the closed-loop system. This is achieved by restricting the supply of energy to the flow's spatial modes. This paper also aims to analyse the linear dynamics of channel flow using the framework of passivity, revealing which modes are responsible for the majority of energy production and therefore which modes need to be controlled. Identifying these spatial modes is a novel contribution and it is hoped that this will aid in future controller design. The work presented in this paper builds on that by Heins et al.~\cite{Heins14} which outlined a linear analysis of this control method. This paper goes substantially further by applying passivity-based controllers to high-fidelity nonlinear simulations in order to gain insight into the performance of this control method on realistic turbulent flows. Finally, the synthesis methodology has been significantly simplified compared to \cite{Sharma11}. We hope that the new synthesis method will be easier to generalise to complex geometries using matrix-free methods that are currently being developed~\cite{Theofilis11}.\\
\\
In the following, a brief review of passivity and passivity-based control is outlined in section~\ref{Passivity}, details of linear analysis and nonlinear numerical testing and a discussion of the new results from this testing is presented in section~\ref{Testing}, conclusions are given in section~\ref{Conclusions}.

\section{Passivity-Based Control}\label{Passivity}

\subsection{Preliminaries}
\begin{figure}
\centering
\includegraphics[width=0.3\textwidth]{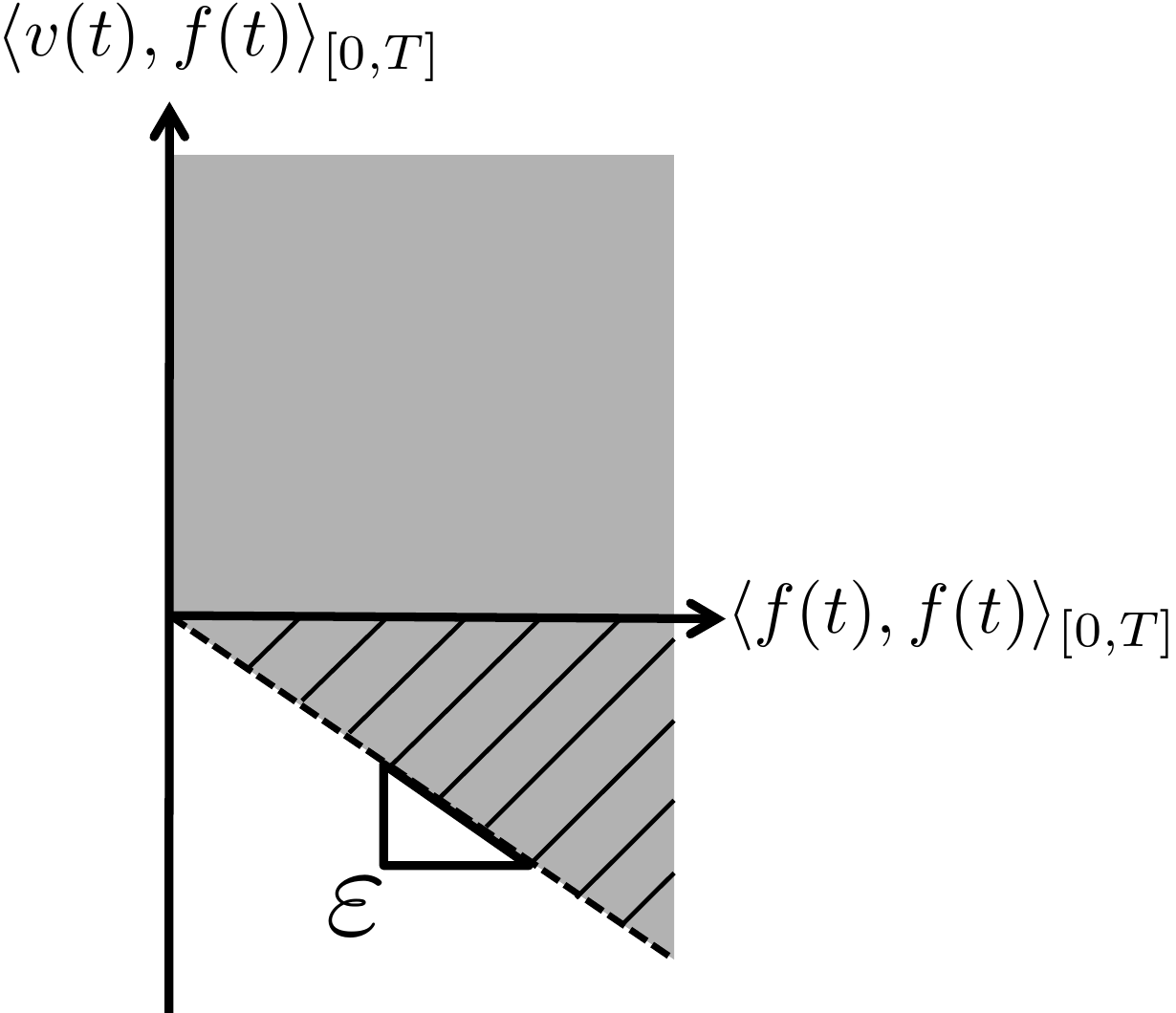}
\caption{A schematic displaying the system energy bound $\varepsilon$ for LTI system $v(t)=Gf(t)$. }
\label{sector}
\end{figure}

Denote with $G$ a LTI representation of a spatially discrete, linearised flow system with the input-output relation $v(t)~=~Gf(t)$, where $v(t)\in\mathbb{R}^{n}$ is a vector of velocities and $f(t)\in\mathbb{R}^n$ is a vector of input forces. The system $G$ is passive if it is only capable of storing and dissipating energy and not producing any of its own. There are several types of passivity and the terminology for each has varied over the years. The conventions of~\cite{Kottenstette14} shall be used in the current work. The system $G$ is \emph{strictly input passive} (SIP) iff there exists $\varepsilon>0$ such that: 
\begin{equation}\label{P1}
\langle v(t),f(t)\rangle_{[0,T]} \geq \varepsilon \langle f(t),f(t) \rangle_{[0,T]} - \Gamma_{0},
\end{equation}
for all $T>0$, where: $\langle X,Y \rangle_{[t_1,t_2]} := \int_{t_1}^{t_2} X^\top Ydt$, denotes an inner product and $\Gamma_{0}\in\mathbb{R}$ is the initial stored energy. The system constant $\varepsilon\in\mathbb{R}$ acts as an energy bound; its importance to the current work will be demonstrated throughout this paper. A system is said to be \emph{passive} if $\varepsilon=0$. For a SIP system, $\varepsilon$ bounds energy dissipation from below. However, if $\varepsilon<0$, the system is not passive and $\varepsilon$ bounds energy production from above. This is demonstrated in the schematic in Fig.~\ref{sector}, for the case where $\Gamma_0=0$.
The graph for $G$ can only lie in the shaded region of the figure. When $\varepsilon<0$, the system is able to dissipate, store \emph{and} produce energy, energy production occurring in the dashed area below the abscissa. The aim of passivity-based control is to minimise $|\varepsilon|$; if possible, forcing $\varepsilon\geq 0$. Note, that minimising the energy bound $|\varepsilon|$ will give no guarantees of either local or global stability. However, it will guarantee a restriction on system energy production from disturbance inputs $f(t)$.   \\
\\
Passivity is a time-domain concept. However, it has a frequency-domain counterpart named \emph{positive realness}. Taking Laplace transforms of system inputs $f(t)$ and outputs $v(t)$, we can form a transfer function matrix $G(s)$ such that $V(s) = G(s)F(s)$ where $s=\sigma + j\omega$ and $j=\sqrt{-1}$. A system with transfer function matrix $G(s)$ is \emph{strictly positive real} (SPR)~\cite{Sun94} iff there exists $\varepsilon > 0$ such that $\forall \omega \in [0,\infty)$:
\begin{equation}
\frac{1}{2} \left [ G(j\omega) + G^\top(-j\omega) \right ] \geq \varepsilon I,
\end{equation}
where $I\in\mathbb{R}^{n\times n}$ is an identity matrix. A system with a SPR transfer function will be SIP~\cite{DesoerBook}. The bound $\varepsilon$ acts in the same manner whether in the time or frequency domain. Therefore, as will be demonstrated in section~\ref{synth}, we can design controllers in the frequency-domain that cause the closed-loop transfer function to be as close to positive real as is possible. This will then mean that in the time-domain our closed-loop system will be as close to passive as is possible. In both cases, this implies minimising the energy bound $|\varepsilon|$.

\subsection{Application to Channel Flow}\label{apply}
\begin{figure}
\centering
\includegraphics[width=0.4\textwidth]{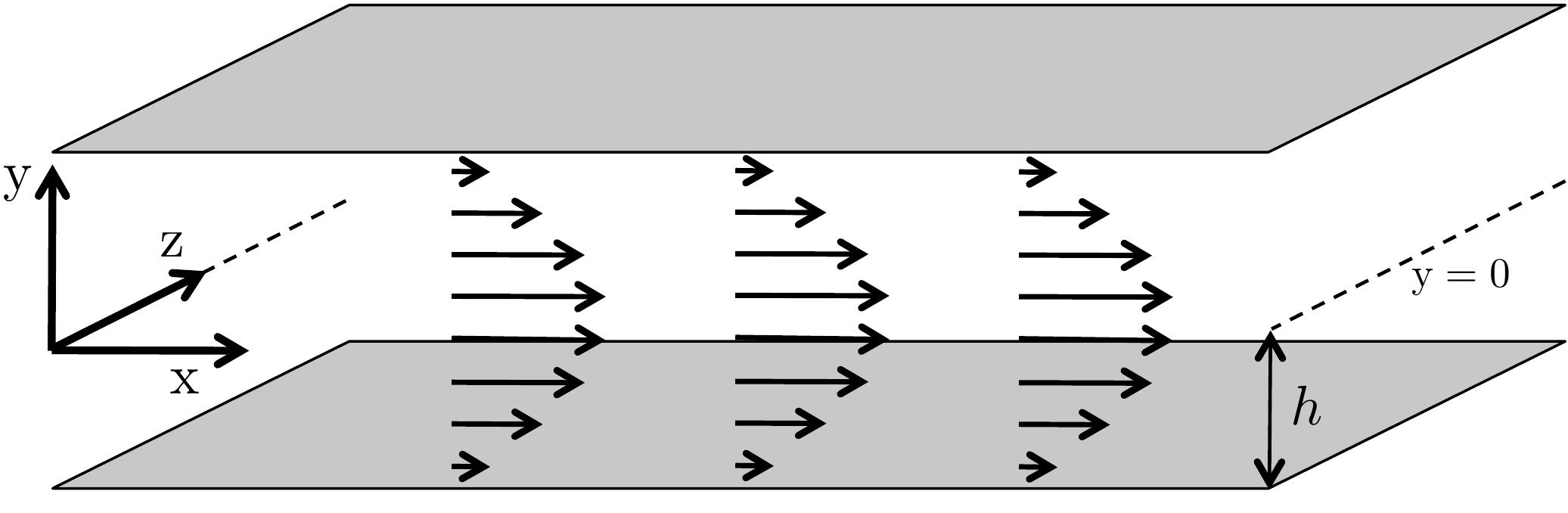}
\caption{A schematic of channel flow. }
\label{channel}
\end{figure}

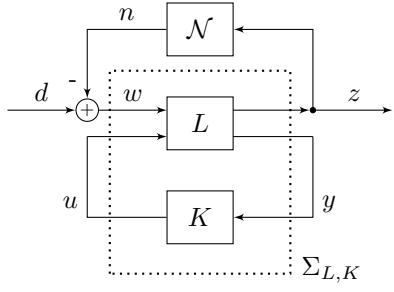
\begin{figure}
\centering
\begin{tikzpicture}[auto, node distance=1.5cm,>=latex']
        \draw [thick, dotted] (-1.2cm,-2cm) rectangle (1.2cm,0.7cm);
    \node [block] (G) {$L$};
    \node [sum, left of=G, yshift=0.5em] (sum) {};
    \node [left of=G, yshift=1.5em, xshift=-0.55em] (minus) {-};
    \node [input, right of=sum, xshift=-1.25em] (Gin1) {};
    \node [input, left of=sum, xshift=1.25em] (disturbance) {};
    \node [block, below of=G, yshift=0.25cm] (controller) {$K$};
    \node [block, above of=G, yshift=-0.25cm] (nonlinearity) {$\mathcal{N}$};
    \node [node, right of=G, yshift=0.5em] (flowfield) {};
    \node [output, left of=flowfield, xshift=1.25em] (Gout1) {};
    \node [output, right of=flowfield, xshift=-1.25em] (out) {};
    \node [output, left of=G, yshift=-0.5em] (waypointl) {};
    \node [input, right of=waypointl, xshift=-1.25em] (Gin2) {};
    \node [output, right of=G, yshift=-0.5em] (waypointr) {};
    \node [input, left of=waypointr, xshift=1.25em] (Gout2) {};
    \node [below of=controller, xshift=1.75cm, yshift=0.8cm] {$\Sigma_{L,K}$};
        \draw [->] (sum) -- node {$w$} (Gin1);
    \draw [->] (Gout1) -- node[near start] {} (flowfield);
    \draw [->] (flowfield) -- node {$z$} (out);
    \draw [->] (disturbance) -- node {$d$} (sum);
    \draw [->] (flowfield) |- node {} (nonlinearity);
    \draw [->] (nonlinearity) -| node[near start, above] {$n$} (sum);
    \draw [-] (controller) -| node {} (waypointl);
    \draw [->] (waypointl) |- node[left, yshift=-2.5em] {$u$} (Gin2);
    \draw [-] (Gout2) -- node {} (waypointr);
    \draw [->] (waypointr) |- node [right, yshift=0.5em] {$y$} (controller);
\end{tikzpicture}
\caption{Block diagram consisting of linear channel flow system $L$ in feedback with nonlinear operator $\mathcal{N}$ and controller $K$.}
\label{Block1}
\end{figure}

Channel flow is the uni-directional flow between two flat plates of infinite spatial dimensions. A schematic of channel flow is shown in Fig.~\ref{channel}. 
The flow is periodic in the streamwise ($\mathrm{x}$) and spanwise ($\mathrm{z}$) directions and inhomogeneous in the wall-normal ($\mathrm{y}$) direction. To avoid notational ambiguity, mathematical descriptions of fluid dynamics will use $\mathrm{Roman}$ font ($\mathrm{x,y,z}$), and $italic$ font shall be used for mathematical descriptions of control systems ($x,y,z$).   \\ 
\\
Incompressible channel flow is governed by the incompressible Navier-Stokes equations~\cite{Aamo} which are a set of PDAEs in terms of velocity $\mathbf{V}$ and pressure $\mathrm{P}$\footnote{All flow variables are non-dimensionalised by maximum laminar centreline velocity $\mathrm{U}_{cl}$ and channel half-height $h$.}. The velocity vector can be decomposed into the summation of a time-independent base flow $\mathbf{\tilde{V}}(\mathrm{y})$ and time-dependent perturbations to the base flow $\mathbf{v}(\mathrm{x},\mathrm{y},\mathrm{z},t)$, such that:
\begin{equation}
\mathbf{V}(\mathrm{x},\mathrm{y},\mathrm{z},t) = \mathbf{\tilde{V}}(\mathrm{y}) + \mathbf{v}(\mathrm{x},\mathrm{y},\mathrm{z},t).
\end{equation}
$\mathbf{v}$ is a vector of the streamwise ($\mathrm{u}$), wall-normal ($\mathrm{v}$) and spanwise ($\mathrm{w}$) perturbation velocities. The pressure field can be decomposed similarly, such that $\mathrm{p}$ denotes perturbation pressure. $\tilde{\mathbf{V}}(\mathrm{y})$ is chosen as an equilibrium solution which is stable to infinitesimal perturbations when $Re<5772$~\cite{Trefethen93}. For a channel flow with walls located at $\mathrm{y}=\pm1$, it is given as  $\tilde{\mathrm{U}} = 1 - \mathrm{y}^2$, $\tilde{\mathrm{V}}=\mathrm{\tilde{W}}=0$. Subtracting $\tilde{\mathbf{V}}(\mathrm{y})$ from the Navier-Stokes equations results in the perturbation equations for a channel flow:
\begin{subequations}
\begin{equation}
\frac{\partial \mathbf{v}}{\partial t} = \frac{1}{Re} \nabla^2 \mathbf{v} - \mathbf{v}\frac{\partial \tilde{\mathrm{U}}}{\partial \mathrm{y}} \mathbf{e}_\mathrm{x} - \tilde{\mathrm{U}} \frac{\partial \mathbf{v}}{\partial \mathrm{x}} - \nabla \mathrm{p} +  \mathbf{d} - \mathbf{n},
\label{P1}
\end{equation}
\begin{equation}
\nabla \cdot \mathbf{v} = 0,
\label{P2}
\end{equation}
\label{NS2}
\end{subequations}
with initial and boundary conditions:
\begin{subequations}
\begin{equation}
\mathbf{v}(\chi,0) = \mathbf{v}_0(\chi) \quad \forall \chi \in \Omega
\end{equation}
\begin{equation}
\mathbf{v}(\chi,t) = \mathbf{g}(\chi,t) \quad \forall (\chi,t) \in \partial \Omega \times [0,t_f],
\end{equation}
\end{subequations}
where $\mathbf{v} : \Omega \times \mathbb{R}_+ \to \mathbb{R}^3$ is the perturbation velocity vector, $\mathrm{p} : \Omega \times \mathbb{R}_+ \to \mathbb{R}$ is the perturbation pressure scalar field, $\mathbf{e}_\mathrm{x}$ is a unit vector pointing in the streamwise direction, $\mathbf{d}: \Omega \times \mathbb{R}_+ \to \mathbb{R}^3$ is a vector of external forces (e.g. exogenous noise), $\mathbf{g}:\partial \Omega \times \mathbb{R}_+ \to \mathbb{R}^3$ is a vector of boundary conditions, $\mathbf{v}_0 \in \mathbb{R}^3$ is an initial velocity vector for $t=0$, $\nabla$ is the gradient operator and $\nabla^2$ is the Laplacian operator. The endpoint of the time interval is $t_f\in\mathbb{R}_+$, $\Omega \subset \mathbb{R}^3$ is a domain in three spatial dimensions with boundary $\partial \Omega$, and $\chi \in \Omega$ is a point within the domain. \\
\\
The nonlinearity can be viewed as a negative feedback forcing upon the linear dynamics of the flow, i.e. $\mathbf{n} = \mathcal{N}(\mathbf{v})$. Therefore, \eqref{P1} can be re-written as:
\begin{equation}
\frac{\partial \mathbf{v}}{\partial t} = \mathbb{L}\left(\mathbf{v},\mathrm{p}\right) + \mathbf{d} - \mathcal{N}\left ( \mathbf{v} \right ),
\end{equation}
where $\mathbb{L}$ denotes an operator comprised of the linear terms in~\eqref{P1}. A schematic of this decomposition is shown in Fig.~\ref{Block1}, where $z(\mathbf{v})$ is an energy-weighted velocity vector and $w(\mathbf{d}-\mathbf{n})$ is an energy-weighted disturbance forcing input to the linear system; both are defined in section~\ref{CMF}. 
For a closed domain such as a channel flow, Sharma et al.~\cite{Sharma11} demonstrated that the nonlinearity is losless, i.e. for all $T>0$:
\begin{equation}
\langle \mathcal{N}\left(\mathbf{v}\right), \mathbf{v}  \rangle_{[0,T]} = 0.
\end{equation}
Therefore, according to the passivity theorem, one is only required to make the feedback system comprising the linear system $L$ and controller $K$, denoted by $\Sigma_{L,K}$ in Fig.~\ref{Block1}, SIP to achieve global stability. This was achieved by Sharma et al.~\cite{Sharma11}. However, this is not possible when actuation is restricted to the wall~\cite{Martinelli11}. Therefore, we aim to use passivity-based control to minimise the energy bound $|\varepsilon_{L,K}|$ of the closed-loop system $\Sigma_{L,K}$. This will not guarantee monotonic decay of perturbation energy of the flow or guarantee local or global stability, but it will restrict the amount of perturbation energy produced by exogenous disturbances $w$. A reduction in perturbation energy causes a reduction in the turbulent fluctuations which in turn leads to a reduction in skin-friction drag~\cite{BewleyMoin01,Sharma11}.

\subsection{Control Model Formulation}\label{CMF}
In order to generate passivity-based controllers, a state-space model of the linear dynamics of channel flow is required. The perturbation equations from \eqref{NS2} in their current form represent a descriptor system due to the divergence-free condition in \eqref{P2} and the pressure term in \eqref{P1}. However, the pressure term can be eliminated and the divergence-free condition satisfied implicitly by transforming the perturbation equations so that they are in terms of only wall-normal velocity ($\mathrm{v}$) and wall-normal vorticity ($\eta_{\mathrm{y}}$)~\cite{SchmidHenningson}. Vorticity is defined as the curl of the velocity vector: $\boldsymbol{\eta} := \nabla \times \mathbf{v}$. In the fluid dynamics literature, the linear perturbation equations for a channel flow in this form are known as the Orr-Sommerfeld Squire equations. As channel flow is periodic in the streamwise ($\mathrm{x}$) and spanwise ($\mathrm{z}$) directions, Fourier spectral discretisation is used in these directions, resulting in wall-normal velocity and vorticity being approximated as:
\begin{subequations}
\begin{equation}
\mathrm{v}(\mathrm{x},\mathrm{y},\mathrm{z},t) \approx \sum_{k_\mathrm{x} = -N_\mathrm{x}/2}^{N_{\mathrm{x}}/2}  \sum_{k_\mathrm{z} = -N_\mathrm{z}/2}^{N_\mathrm{z}/2} \hat{\mathrm{v}}_{\alpha , \beta}(\mathrm{y},t)e^{j(\alpha \mathrm{x} + \beta \mathrm{z})},
\end{equation}
\begin{equation}
\eta_{\mathrm{y}}(\mathrm{x},\mathrm{y},\mathrm{z},t) \approx \sum_{k_\mathrm{x} = -N_\mathrm{x}/2}^{N_{\mathrm{x}}/2}  \sum_{k_\mathrm{z} = -N_\mathrm{z}/2}^{N_\mathrm{z}/2} \hat{\eta}_{\mathrm{y},\alpha , \beta}(\mathrm{y},t)e^{j(\alpha \mathrm{x} + \beta \mathrm{z})},
\end{equation}
\end{subequations}
where $\hat{\mathrm{v}}_{\alpha,\beta}\in\mathbb{C}$ and $\hat{\eta}_{\mathrm{y},\alpha,\beta}\in\mathbb{C}$ are Fourier spectral coefficients, $N_{\mathrm{x}}\in\mathbb{Z}$ and $N_{\mathrm{z}}\in\mathbb{Z}$ are the number of discrete Fourier modes used in the streamwise and spanwise spatial directions respectively, $\alpha := 2\pi k_\mathrm{x}/L_\mathrm{x}\in\mathbb{R}$ and $\beta:= 2\pi k_\mathrm{z}/L_\mathrm{z}\in\mathbb{R}$ are the streamwise and spanwise wavenumbers respectively, and $L_\mathrm{x},L_\mathrm{z} \in \mathbb{R}$ are the streamwise and spanwise dimensions of the channel. This results in spatially one-dimensional models for each wavenumber pair ($\alpha,\beta$) that one wishes to control, the dynamics of which are: 
\begin{equation}
\underbrace{\frac{d}{dt} \left [ \begin{array}{c}  \hat{\mathrm{v}}(\mathrm{y},t) \\ \hat{\eta_{\mathrm{y}}}(\mathrm{y},t) \end{array} \right ]}_{\displaystyle \dot{x}} \underbrace{\left [ \begin{array}{cc} \mathbb{L}_{OS} & 0 \\ \mathbb{L}_C & \mathbb{L}_{Sq} \end{array} \right ]}_{\displaystyle A} \underbrace{\left [ \begin{array}{c} \hat{\mathrm{v}}(\mathrm{y},t) \\ \hat{\eta_{\mathrm{y}}}(\mathrm{y},t) \end{array} \right ]}_{\displaystyle x} ,
\end{equation}  
where $\mathbb{L}_{OS}$, $\mathbb{L}_{C}$ and $\mathbb{L}_{Sq}$ are the Orr-Sommerfeld, coupling and Squire operators respectively~\cite{SchmidHenningson,schmid2007}, the state vector $x\in\mathbb{C}^{p1}$ is chosen to be the wall-normal velocity and vorticity spectral coefficients, and dot notation represents a derivative with respect to time. A Chebyshev collocation method~\cite{TrefethenSPM} is used to discretise the inhomogenous wall-normal ($\mathrm{y}$) direction. The method used ensures that the following boundary conditions on $\hat{\mathrm{v}}$ and $\hat{\eta_{\mathrm{y}}}$ are satisfied:
\begin{subequations}
\begin{equation}
\frac{d\hat{\mathrm{v}}(\pm1,t)}{dt} = -\frac{1}{\tau_\phi}\hat{\mathrm{v}}(\pm1,t) + \frac{1}{\tau_\phi}q_\pm,
\end{equation}
\begin{equation}
\frac{\partial \hat{\mathrm{v}}(\pm1,t)}{\partial \mathrm{y}} = \hat{\eta}_\mathrm{y}(\pm1,t) = 0,
\end{equation}
\end{subequations}
with initial boundary conditions on $\hat{\mathrm{v}}$:
\begin{equation}
\hat{\mathrm{v}}(\pm1,0) = 0,
\end{equation}
where a low-pass filter has been applied to the actuation dynamics as previously done by~\cite{Jones15}, $\tau_\phi$ is the actuator time constant, and $q_+$ and $q_-$ are the control signals to the upper and lower walls respectively. A lifting procedure~\cite{McKernan06} is used to incorporate actuation.\\
\\
The chosen measurement output vector $y\in\mathbb{C}^{q_2}$ consists of the streamwise ($\hat{\tau}_{\mathrm{y}\mathrm{x}}$) and spanwise ($\hat{\tau}_{\mathrm{y}\mathrm{z}}$) wall-shear stresses at each wall. These are defined as:
\begin{equation}
y = \left [ \begin{array}{c} \hat{\tau}_{\mathrm{y}\mathrm{x}} |_{\mathrm{y}=\pm1} \\ \hat{\tau}_{\mathrm{y}\mathrm{z}} |_{\mathrm{y}=\pm1} \end{array} \right ]
= \frac{1}{Re} \left [ \begin{array}{c} \left ( \frac{\partial \hat{\mathrm{u}}}{\partial \mathrm{y}} +  \frac{\partial \hat{\mathrm{v}}}{\partial \mathrm{x}} \right ) |_{\mathrm{y}=\pm1} \\ \left ( \frac{\partial \hat{\mathrm{w}}}{\partial \mathrm{y}}  +  \frac{\partial \hat{\mathrm{v}}}{\partial \mathrm{z}} \right )|_{\mathrm{y}=\pm1}  \end{array} \right ] .
\label{y1}
\end{equation}  
Using \eqref{P2} and the definition of vorticity, expressions for $\hat{\mathrm{u}}$ and $\hat{\mathrm{w}}$ in terms of $\hat{\mathrm{v}}$ and $\hat{\eta_{\mathrm{y}}}$ can be formed which in turn are used to transform~\eqref{y1} to be in terms of the state variables. \\
\\
The controlled output $z\in\mathbb{C}^{q_1}$ is required to be an energy weighted vector of the states, such that perturbation energy $E \approx z^\top z \in \mathbb{R}$. Perturbation energy is defined as~\cite{Bewley98}:
\begin{equation}
E := \frac{1}{8} \int_{-1}^1 \hat{\mathrm{v}}^\top \hat{\mathrm{v}} + \frac{1}{k^2}\left ( \frac{\partial \hat{\mathrm{v}}}{\partial \mathrm{y}}^\top  \frac{\partial \hat{\mathrm{v}}}{\partial \mathrm{y}} + \hat{\eta_{\mathrm{y}}}^\top\hat{\eta_{\mathrm{y}}}  \right ) d \mathrm{y},
\label{E1}
\end{equation}
where $k^2 := \alpha^2 + \beta^2$. Using Chebyshev differentiation matrices~\cite{Weideman00} to perform the discrete differentiation and Clenshaw-Curtis quadrature~\cite{TrefethenSPM} to perform the discrete integration in~\eqref{E1}, $E$ can be approximated as:
\begin{equation}
E \approx x^\top Q x = (C_1x)^\top C_1x,
\end{equation}
where $Q$ is a discrete energy matrix and $C_1$ can be found by performing a Cholesky decomposition on $Q$.\\
\\
The preliminary state-space model $L$ for the flow at each wavenumber pair ($\alpha,\beta$) is represented as:
\begin{equation}
\left [ \begin{array}{c} \dot{x} \\ z \\ y \end{array} \right ] = \underbrace{\left [ \begin{array}{c|cc} A & B_1 & B_2 \\
																\hline
															             C_1 & 0 & 0\\
															             C_2 & 0 & 0 \end{array} \right ]}_{\displaystyle L} \left [ \begin{array}{c}  x \\ w \\ u \end{array} \right ] 
															             \label{W1}
\end{equation}
where $B_1 = C_1^{-1}$ in order for exogenous disturbance inputs $w\in\mathbb{C}^{r_1}$ to be energy weighted appropriately, and control input vector $u= [ \begin{array}{cc} q_+ & q_- \end{array}  ]^\top \in \mathbb{C}^{r_2}$. The model needs to be augmented to include a feed-through term $D_{11} = \bar{\varepsilon}I$ which models feed-through energy $\bar{\varepsilon}>0 \in \mathbb{R}$. Feed-through energy is the primary tuning constant when designing passivity-based controllers. Its importance will be made apparent in the following section. The control model then needs to be augmented once more to include penalties on the control signals $\epsilon_c\in\mathbb{R}$ and measurement quality $\epsilon_d\in\mathbb{R}$. This is achieved through the addition of feed-through matrices $D_{12}(\epsilon_c)$ and $D_{21}(\epsilon_d)$ which are given in the final control model $\tilde{L}$ below: 
\begin{equation}
\tilde{L} := \left [ \begin{array}{c|cc} A & \tilde{B}_1 & B_2 \\
							\hline
						\tilde{C}_1 & \tilde{D}_{11} & D_{12} \\
						C_2 & D_{21} & 0 \end{array} \right ] :=  \left [ \begin{array}{c|cc} A &   [ \begin{array}{cc} B_1&0\end{array} ] & B_2 \\
								\hline
							      \left [ \begin{array}{c} C_1 \\ 0 \end{array} \right ] & \left [ \begin{array}{cc} D_{11} & 0 \\ 0 & I \end{array} \right ] & \left [ \begin{array}{c} 0 \\ \epsilon_c I \end{array} \right ]\\
							      C_2 & [ \begin{array}{cc} 0 & \epsilon_d I \end{array} ] & 0 \end{array} \right ]
.							      
\end{equation}

\subsection{Passivity-Based Controller Synthesis}\label{synth}
The procedure used to generate passivity-based controllers is based on that outlined by Sun et al.~\cite{Sun94}; this procedure differs to that used by Sharma et al.~\cite{Sharma11} which required a bilinear transformation to an equivalent $\mathcal{H}_\infty$ problem. They outlined a method of finding LTI controllers that when implemented in feedback with a LTI plant, force the closed-loop system's transfer function to be SPR. As discussed in section~\ref{intro}, a system with a SPR transfer function is also SIP. Controllers are formed from the solutions to two algebraic Riccati equations (AREs) which in turn are functions of all the matrices in our control model. For a given wavenumber pair and sensing/actuation arrangement, solutions to the AREs only exist depending on the values of the tuning constants  $\epsilon_c$, $\epsilon_d$ and primarily $\bar{\varepsilon}$. \\
\\
The final controller $K$ will be of the form:
 \begin{equation}
 \left [ \begin{array}{c} \dot{x}_K \\ u \end{array} \right ] = \left [ \begin{array}{c|c} A_K & B_K \\
 																\hline
															C_K & 0 \end{array} \right ]  \left [ \begin{array}{c} {x}_K \\ y \end{array} \right ],	
 \end{equation}
where $x_K\in\mathbb{C}^{p2}$ are the states of the controller and estimates of the states of $\tilde{L}$. The closed-loop system $\Sigma_{\tilde{L},K}$ comprised of the linear plant $\tilde{L}$ in feedback with the controller $K$ will have the form:
\begin{equation}
\Sigma_{\tilde{L},K} := \left [ \begin{array}{c|c} A_{\tilde{L},K} & B_{\tilde{L},K} \\
						  						\hline
												C_{\tilde{L},K} & \tilde{D}_{11} \end{array} \right ] := \left [ \begin{array}{cc|c} A & B_2C_K & \tilde{B}_1\\
						  B_KC_2 & A_K & B_KD_{21} \\
								  \hline
						  \tilde{C}_1 & D_{12}C_K & \tilde{D}_{11} \end{array} \right ]
						  .
\end{equation}
$K$ is required to ensure two things. Firstly, that $\Sigma_{\tilde{L},K}$ is internally stable. Secondly, that $\Sigma_{\tilde{L},K}$ is SPR. This last requirement may seem at odds with what has been previously stated. However, the control model $\tilde{L}$ includes a feed-through energy term $D_{11}=\bar{\varepsilon}I$ which is artificial and absent from the actual system's dynamics. The artificial feed-through energy included in the control model $\tilde{L}$ acts as the final energy bound when the closed-loop system comprised of $K$ with the actual system $L$ is formed. This will be demonstrated below.\\
\\
Defining the control model closed-loop system's transfer function:
\begin{equation}
\Sigma_{\tilde{L},K}(s) := C_{\tilde{L},K} \left ( sI - A_{\tilde{L},K} \right )^{-1}B_{\tilde{L},K} + \tilde{D}_{11} ,
\end{equation}
such that $Z(s) = \Sigma_{\tilde{L},K}(s) W(s)$, where $Z(s)$ and $W(s)$ are the Laplace-transformed controlled output vector $z(t)$ and disturbance input vector $w(t)$ respectively, the closed-loop system is SPR iff $\forall \omega \in [0,\infty )$:
\begin{equation}
\frac{1}{2} \left ( \Sigma_{\tilde{L},K}(j \omega ) +  \Sigma_{\tilde{L},K}^\top (-j \omega ) \right ) > 0.
\label{In1}
\end{equation}
Taking the feed-through terms to the right hand side of the inequality in~\eqref{In1}:
\begin{equation}
\frac{1}{2} \left ( {\Sigma}_{{L},K}(j \omega ) +  {\Sigma}_{{L},K}^\top (-j \omega ) \right ) > -\bar{\varepsilon}I \approx \varepsilon_{L,K}I,
\end{equation}
where ${\Sigma}_{{L},K} = C_{\tilde{L},K}  ( sI - A_{\tilde{L},K}  )^{-1}B_{\tilde{L},K}$. It can now be seen how the feed-through energy term relates to the final closed-loop energy bound $\varepsilon_{L,K} \approx -\bar{\varepsilon}<0$. Being a suboptimal procedure, equality is only reached for $\bar{\varepsilon}_{\text{min}}$ which is found via iteration. The procedure for generating $K$ for a given wavenumber pair is therefore:
\begin{enumerate}
\item Choose tuning constants $\epsilon_d$ and $\epsilon_c$ both of which should be less than unity and set $\tau_\phi\leq1$.
\item Calculate the open-loop energy bound $\varepsilon_L$ of the uncontrolled system $L$ using \eqref{OLvar}.
\item Set $\bar{\varepsilon} = -\varepsilon_L$, form the control model $\tilde{L}$ and solve the AREs outlined in Sun et al.~\cite[p.~2037]{Sun94}.
\item Reduce $\bar{\varepsilon}$ until stabilising solutions to the AREs no longer exist.
\item The minimal feed-through energy for which solutions exist will equal the tightest closed-loop energy bound achievable when the controller $K$ is formed and then implemented in feedback with linear plant $L$: $\bar{\varepsilon}_{\mathrm{min}} = |\varepsilon_{L,K}|_{\mathrm{min}}$. 
\end{enumerate}

\section{Testing and Results}\label{Testing}
\begin{figure}
\centering
\includegraphics[width=0.32\textwidth]{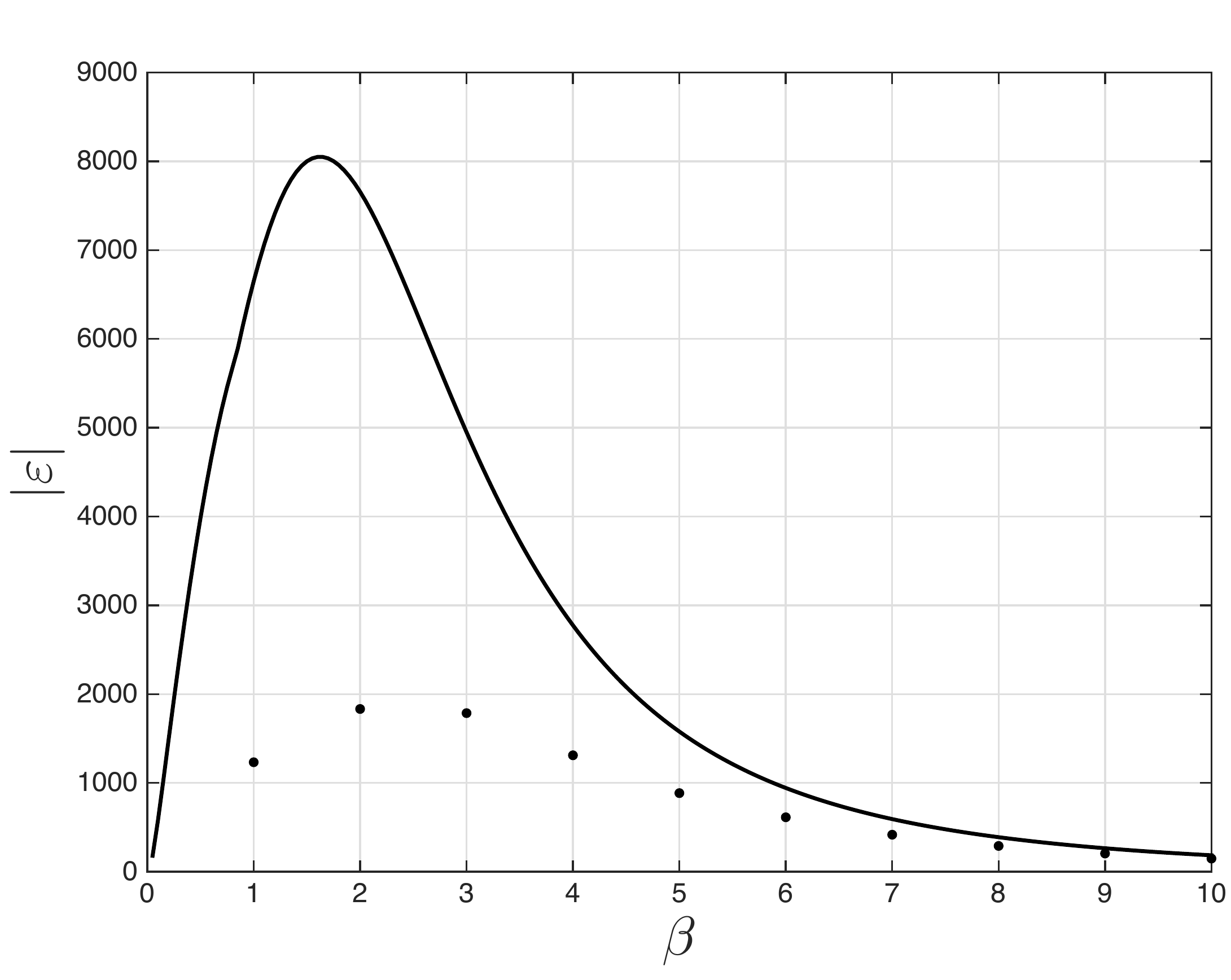}
\caption{Plot of $|\varepsilon_{L}|$ (black line) and $|\varepsilon_{L,K}|$ (black dots) for $\alpha=0,\beta\leq10$, $Re=2230$.}
\label{OLCL}
\end{figure}

All controllers were designed for and tested on $Re=2230$ channel flows. This maximum centreline velocity Reynolds number was used as it corresponds to $Re_\tau=100$ fully turbulent channel flow; this is the same $Re_\tau$ as used by Sharma et al.~\cite{Sharma11} and H\"ogberg et al.~\cite{Hogberg03b}, for which the current work is directly comparable. This Reynolds number is subcritical for a channel flow, meaning that the system is open-loop stable {to infinitesimal perturbations}. However, the eigenvectors of the linear system form a non-orthogonal set. This can result in large transient energy growth of several orders of magnitude relative to the initial condition in some spatial modes~\cite{Butler92,schmid2007,Trefethen93}. Large energy transients above a certain threshold can lead to subcritical transition to turbulence via nonlinear mechanisms. Once a flow is in a turbulent state, the turbulent fluctuations are self-sustaining. \\
\\
Before generating any controllers, it was first deemed necessary to investigate what regions of wavenumber space have the largest open-loop bounds on energy production $\varepsilon_L$. For a given wavenumber pair $(\alpha,\beta)$:
\begin{equation}
\varepsilon_L := \frac{1}{2}\left \{ \inf_\omega \lambda_{\mathrm{min}} \left (L(j\omega) + L^\top(-j\omega) \right) \right \},
\label{OLvar}
\end{equation}
where $L(s) = C_1(s I-A)^{-1}B_1$ is the open-loop transfer function matrix of the system and $\lambda_{\text{min}}$ denotes the smallest eigenvalue.  
\begin{figure}
\centering
\includegraphics[width=0.45\textwidth]{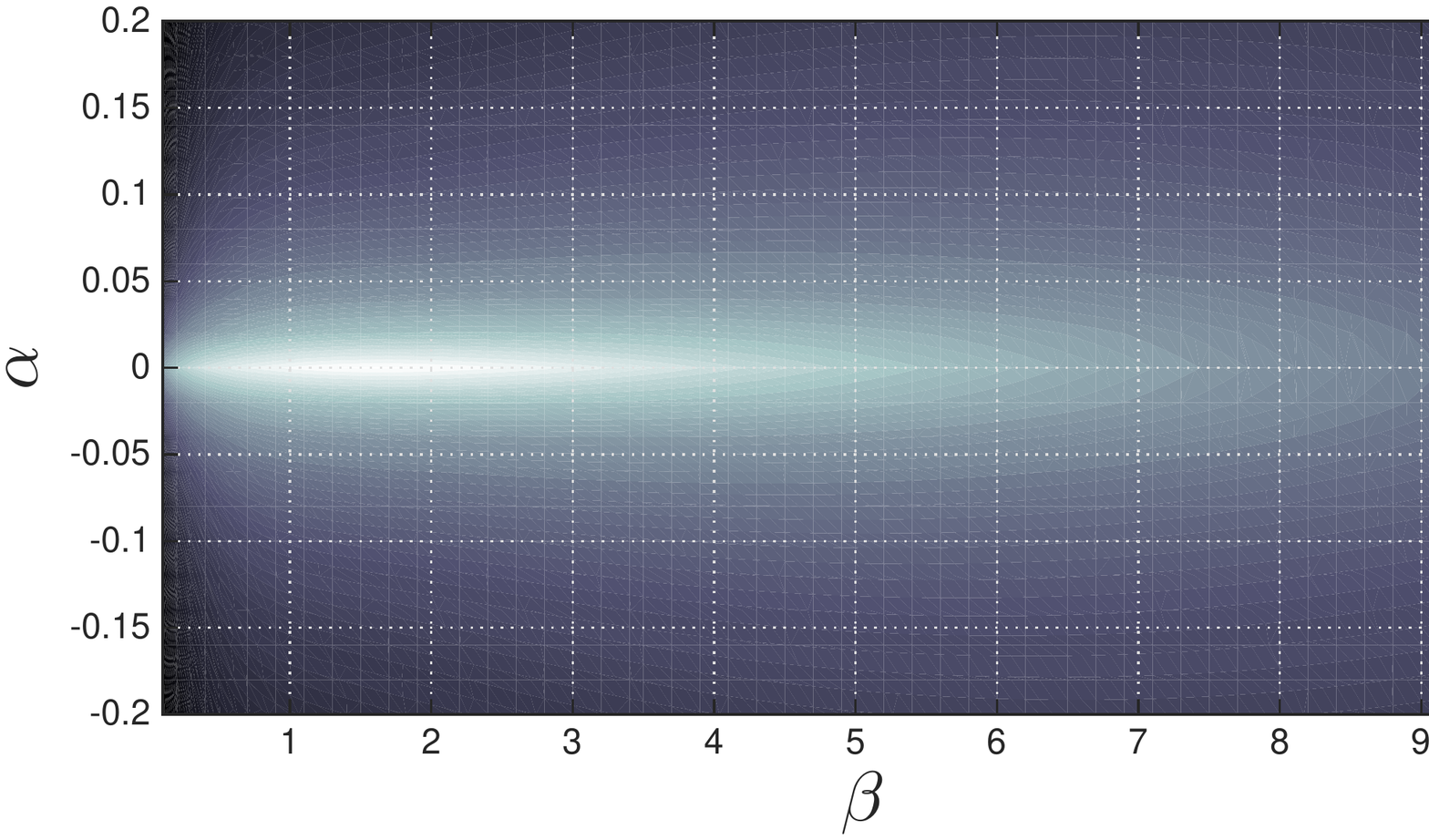}
\caption{Map of $\log_{10}|\varepsilon_L|$ in wavenumber space.}
\label{Vareps_OL}
\end{figure}
Fig.~\ref{Vareps_OL} shows a map of $\log_{10}|\varepsilon_L|$ in wavenumber space. As can be seen, the largest values by far are for the streamwise constant wavenumber pairs $\alpha=0,\beta \neq0$. Wavenumber pairs with non-zero $\alpha$ have $\varepsilon_L$ values that are negligible relative to the streamwise constant modes. Streamwise constant modes where $\beta>10$ have similar $\varepsilon_L$ values as the non-zero $\alpha$ modes. The region of wavenumber space within $\alpha=0,\beta\leq3$ has particularly large  values of  $|\varepsilon_L|\geq5000$, meaning that it is these wavenumber pairs which are capable of producing the most perturbation energy within the flow. The wavenumber pair found to have the highest  $|\varepsilon_L|$ for all Reynolds numbers tested is $\alpha=0,\beta=1.62$. This result compares with the research of Trefethen et al.~\cite{Trefethen93}, who found that it was this wavenumber pair that displayed maximum resonance to exogenous disturbances for a channel flow.    \\
\\
Passivity-based controllers were generated for several streamwise constant wavenumber pairs. Fig.~\ref{OLCL} shows a graph of energy bounds $|\varepsilon_L|$ and $|\varepsilon_{L,K}|$ for wavenumber pairs $\alpha=0,\beta \leq 10$.
As can be seen, the controllers achieve large reductions in $|\varepsilon|$ for the lowest three streamwise constant modes. The controllers for $\alpha=0,\beta=1$ and $\alpha=0,\beta=2$ enforce reductions in $|\varepsilon|$ of $81\%$ and $75\%$ respectively. The reductions in $|\varepsilon|$ due to the controllers tends to zero for $\beta>10$. Therefore, it was decided that controllers would only be needed for the ten lowest streamwise constant modes. The final controllers all had penalties $\epsilon_c = \epsilon_d = 0.001$, filter time constant $\tau_\phi=0.01$ and wall-normal resolution $N_{\mathrm{y},c} = 168$. This $N_{\mathrm{y},c}$ was large enough for solutions to the AREs to converge.  \\
\\
In order to evaluate the robustness 
\begin{figure}
\centering
\includegraphics[width=0.32\textwidth]{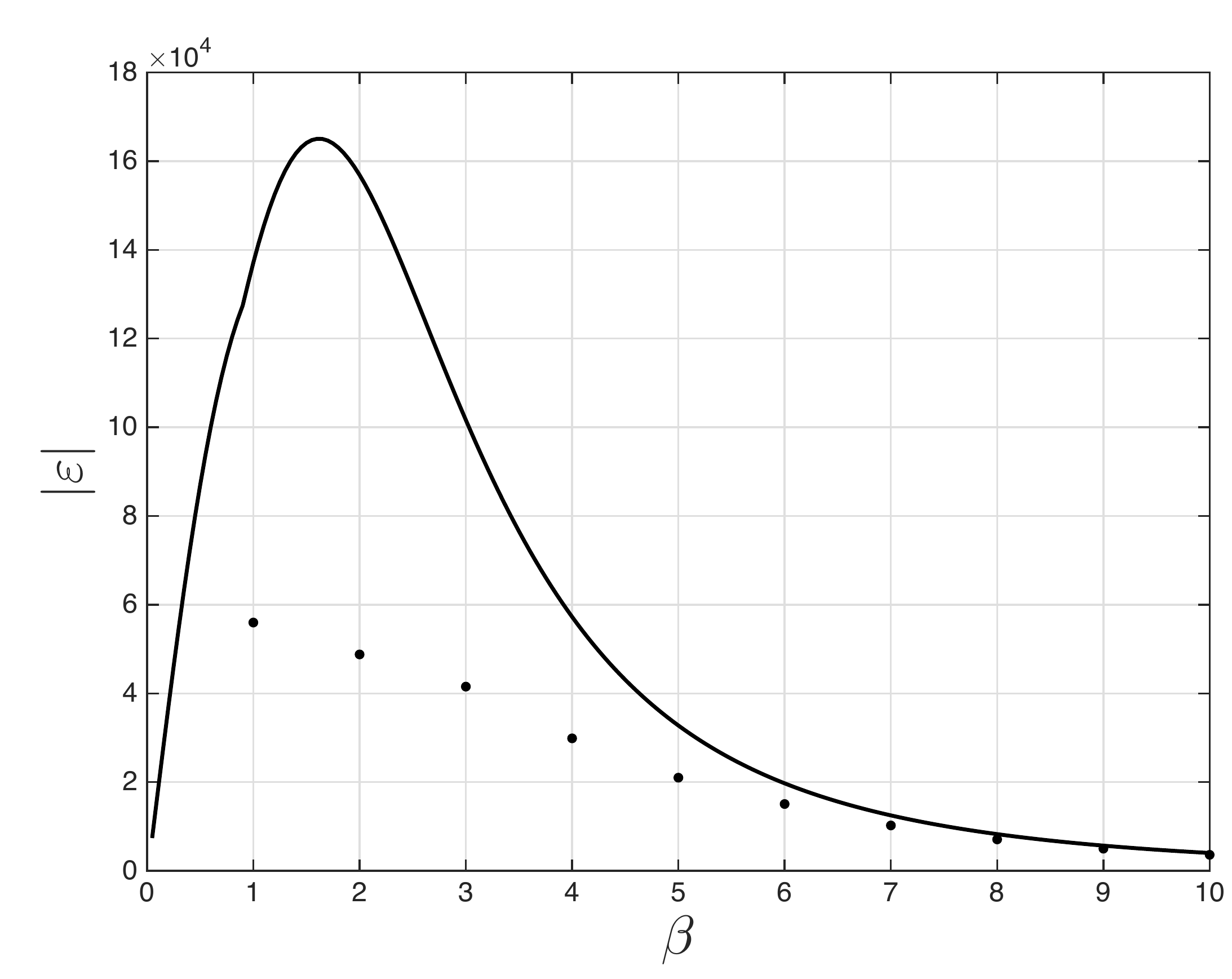}
\caption{Plot of $|\varepsilon_{L}|$ (black line) and $|\varepsilon_{L,K}|$ (black dots) for $\alpha=0,\beta\leq10$, $Re=10^4$.}
\label{robustness}
\end{figure}
of the controllers to variations in Reynolds number, controllers designed for $Re=2230$ channel flow were connected in feedback with higher Reynolds number linear plants corresponding to the ten controlled modes.  
All controllers were locally stabilising to small perturbations up to $Re=10^4$. Fig.~\ref{robustness} shows the variation of $|\varepsilon_{L}|$ and $|\varepsilon_{L,K}|$ for the ten controlled modes for plant $Re=10^4$. The two lowest modes seem to be most affected by the change in plant Reynolds number. Controllers for these modes now only reduce $|\varepsilon|$ by $59\%$ and $69\%$ for $\alpha=0,\beta=1$ and $\alpha=0,\beta=2$ respectively. However, bearing in mind that the Reynolds number of the plant is nearly five times higher than the controller design Reynolds number, these are still significant reductions.

 \subsection{DNS Testing}
\begin{figure}
\begin{center}
\includegraphics[width=0.31\textwidth]{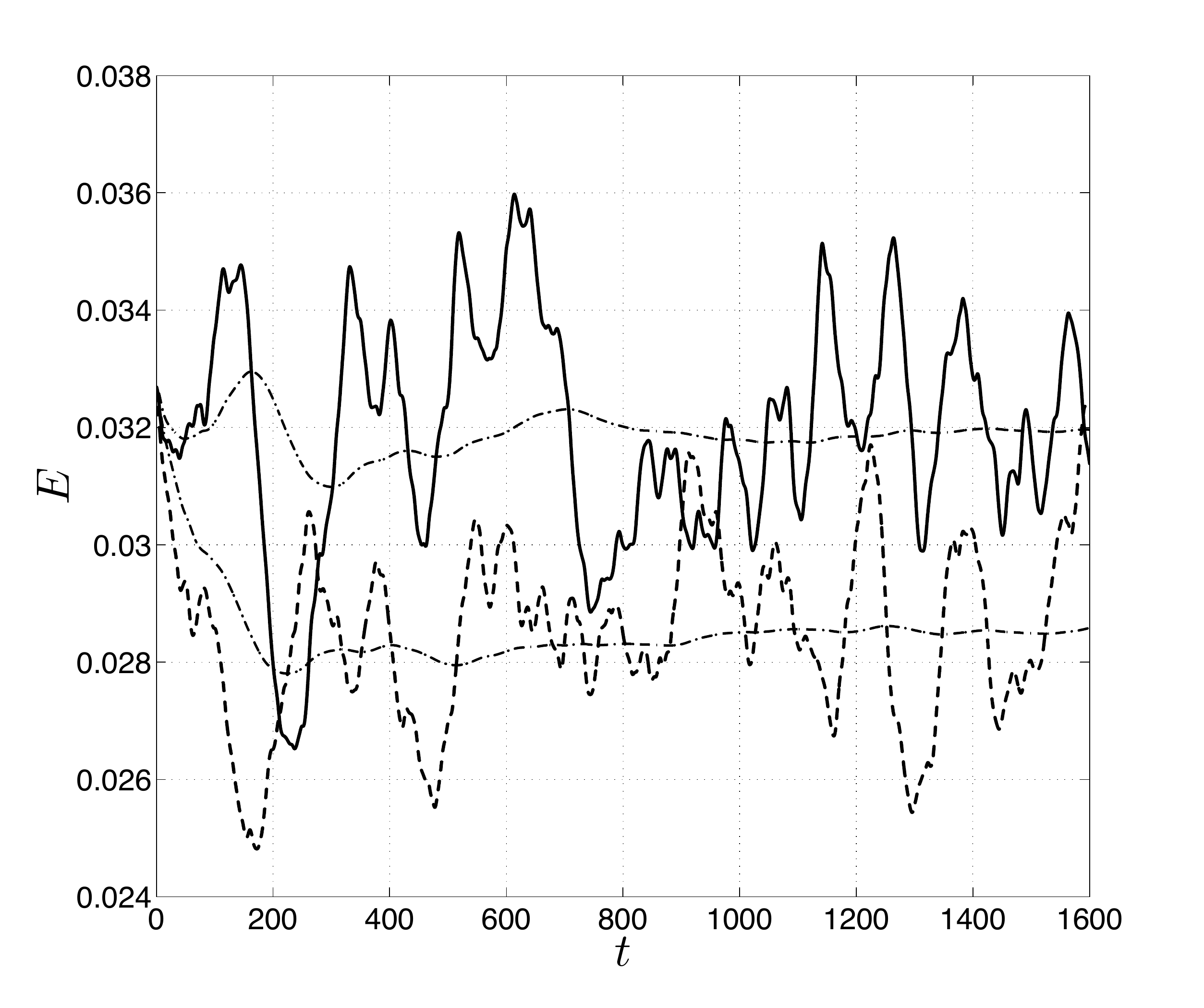}
\caption{Plots of total perturbation energy over time for the uncontrolled (thick line) and controlled (dashed line) simulations. Also included are dash-dot lines representing the uncontrolled and controlled moving mean values.}
\label{energy}
\end{center}
\end{figure}

\begin{figure}
\begin{center}
\includegraphics[width=0.38\textwidth]{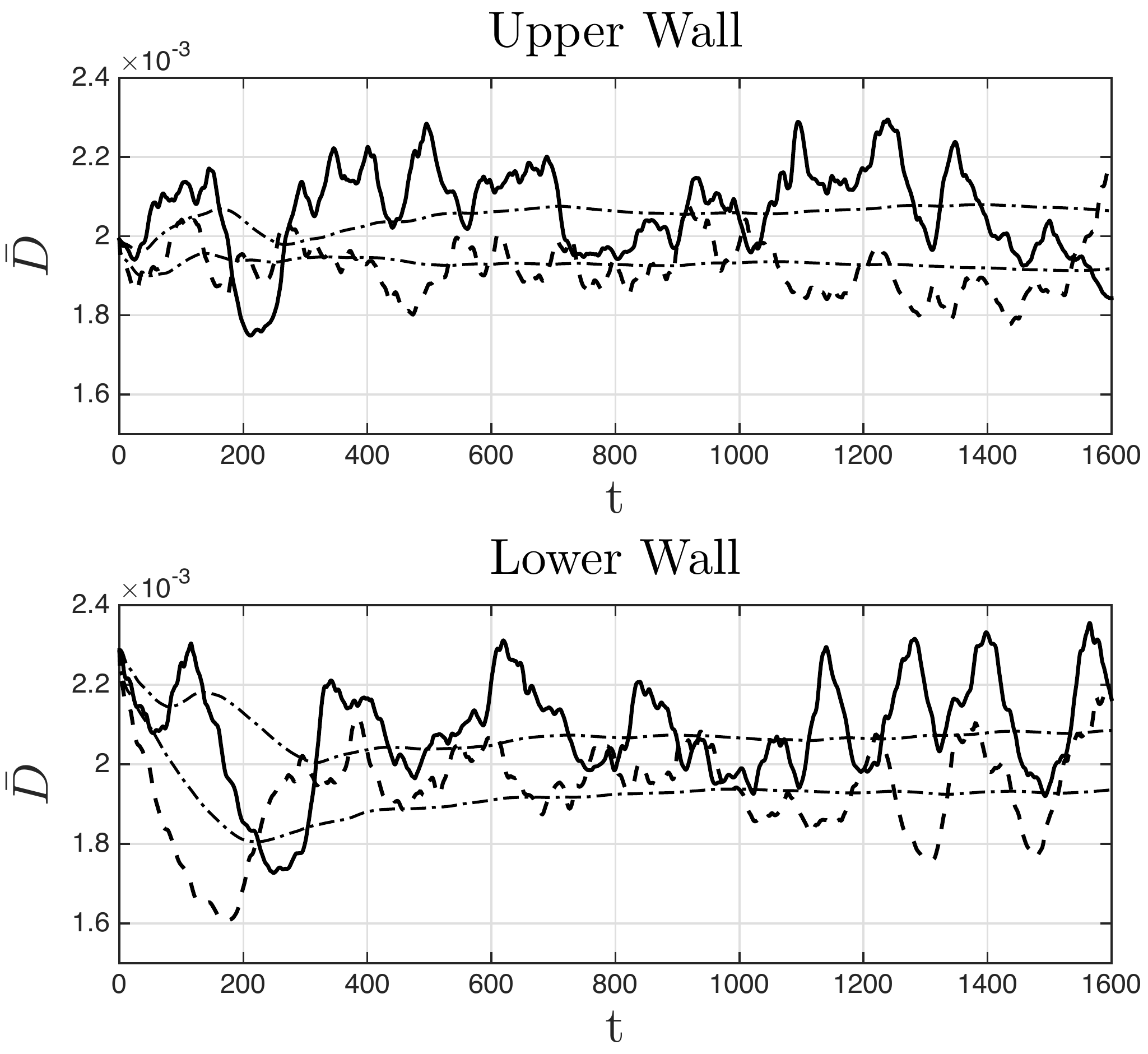}
\caption{Plots of total drag over time for the upper and lower walls for the uncontrolled flow (thick line) and controlled flow (dashed line). Also included are dash-dot lines representing moving mean values. }
\label{drag}
\end{center}
\end{figure}

\begin{figure}
\begin{center}
\includegraphics[width=0.9\columnwidth]{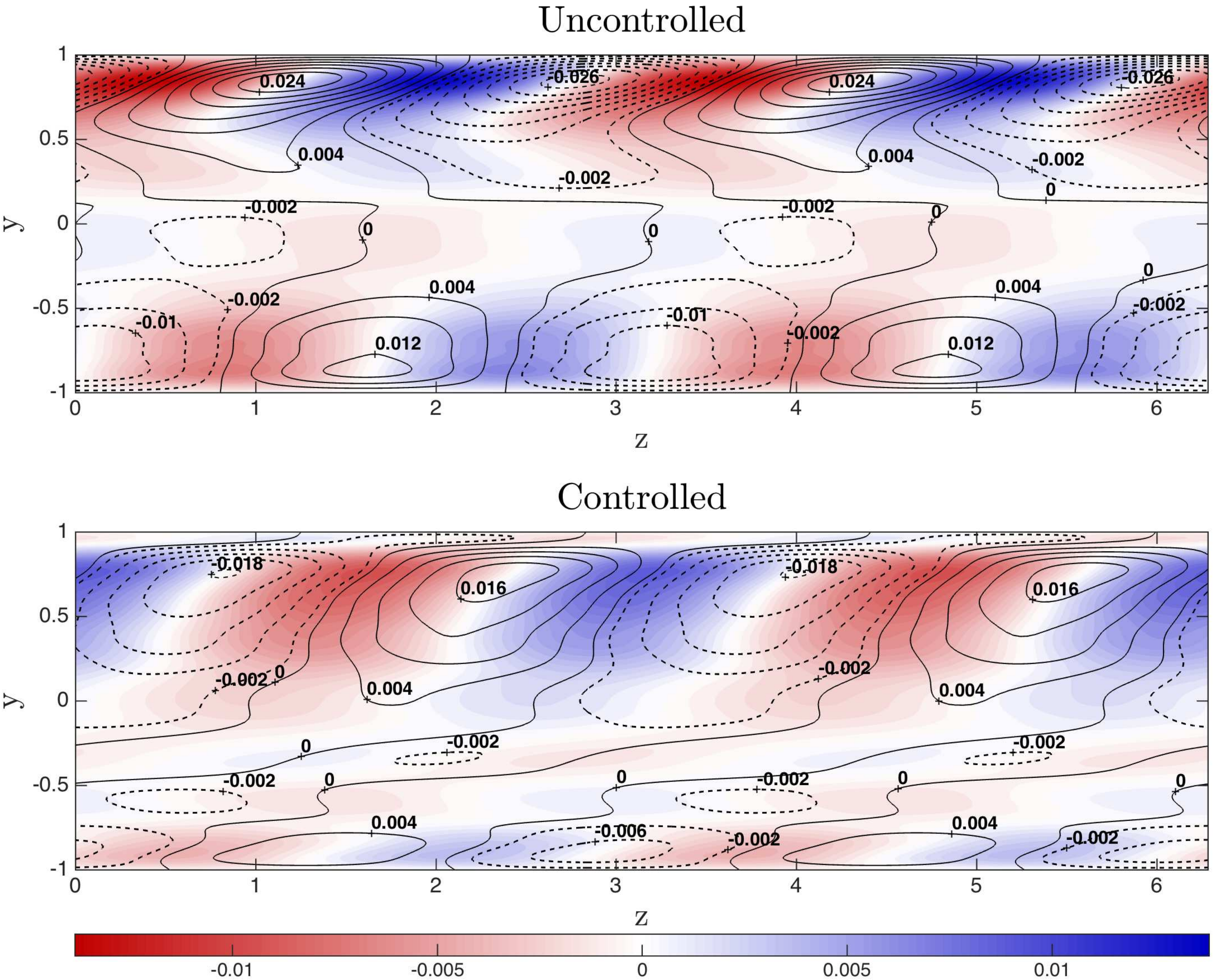}
\caption{Plots of streamwise constant perturbation streamwise velocity (colour map) and perturbation streamwise vorticity (thick contours for positive values and dashed contours for negative values) for wavenumber pair $\alpha=0,\beta=2$ and $T=1100$. }
\label{swvelvort}
\end{center}
\end{figure}

The controllers were evaluated upon a direct numerical simulation (DNS) of $Re_\tau=100$ fully turbulent channel flow. DNS programs integrate the discretised nonlinear incompressible Navier-Stokes equations forward in time from a given initial condition. A modified version of the open-source serial DNS code \emph{Channelflow}~\cite{channelflow} was used in the current work. The program was modified to allow for inhomogeneous boundary conditions to be set at each simulation time step. Channelflow uses a spectral discretisation - Fourier $\times$ Chebsyshev-collocation $\times$ Fourier in the streamwise, wall-normal and spanwise directions respectively. For all testing, the nonlinearity was computed in skew-symmetric form, the simulation was marched forward in time using a 3rd-order semi-implicit backward differentiation algorithm, flow bulk velocity was kept constant and kinematic viscosity was set at $\nu = \frac{1}{Re} = \frac{1}{2230}$. The spatial domain used for all simulations had dimensions $4\pi  \times 2 \times 2\pi $ and resolution $182\times 151\times158$ in the streamwise, wall-normal and spanwise directions respectively. All controlled simulations were started from a fully developed turbulent flow field. The benchmark data of  Iwamoto~\cite{Iwamoto02} was used to check that this flow field was statistically steady-state turbulent. Controlled simulations were run for $1600$ dimensionless time units. This was sufficient for mean values of skin-friction drag and energy to converge. \\
 \\
 Fig.~\ref{energy} shows plots of total perturbation energy over time. 
The combined effort of the controllers resulted in an overall reduction in mean total perturbation energy of $11\%$. As seen from Fig.~\ref{energy}, when the controllers are initially activated, the total perturbation energy of the flow reduces greatly for the first $180$ time units. This is a transient phenomenon and afterwards the perturbation energy recovers but at a lower mean steady-state level than the uncontrolled flow. 
\begin{figure}
\begin{center}
\includegraphics[width=0.31\textwidth]{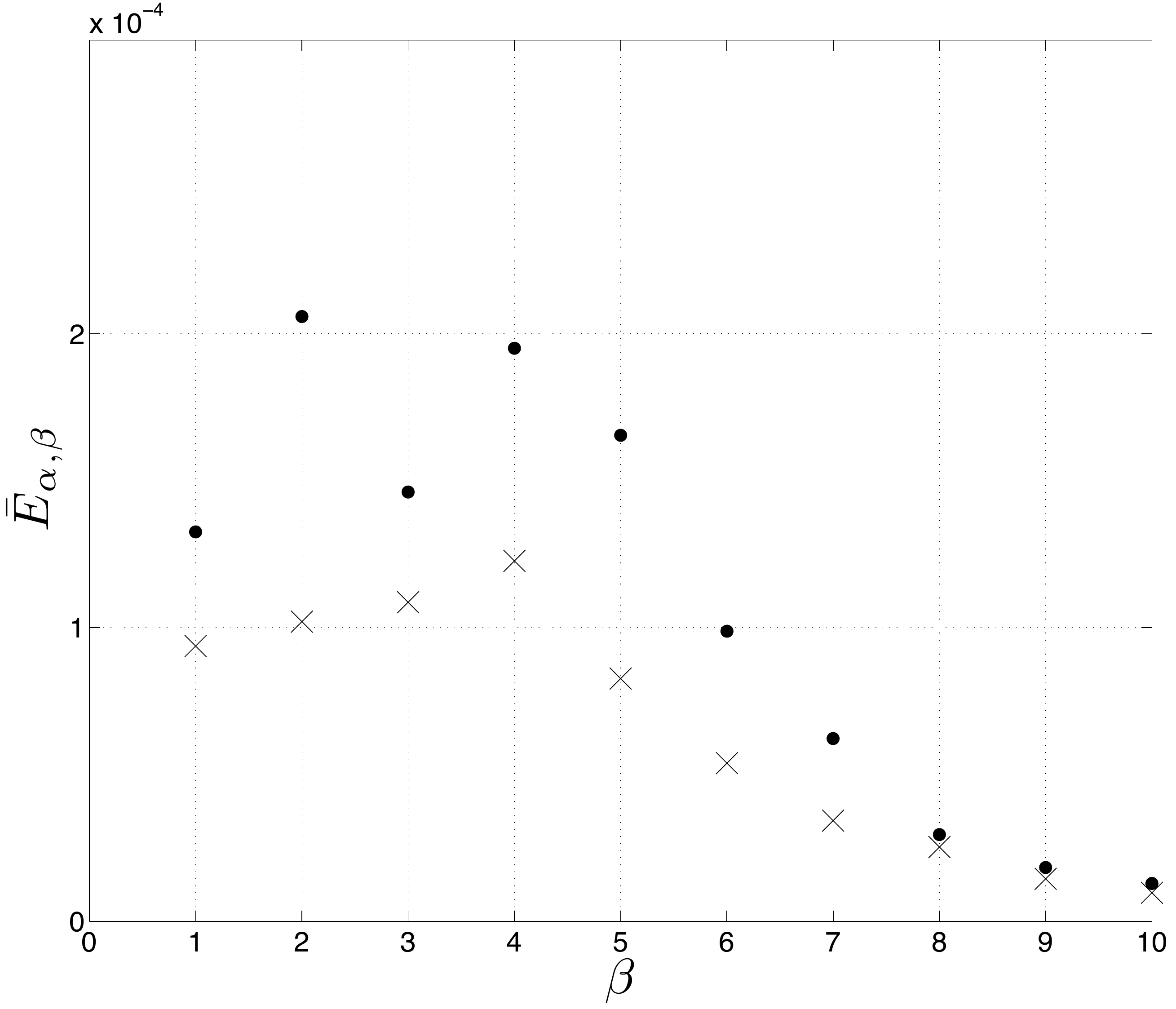}
\caption{Plots of mean energy for modes $\alpha=0,\beta\leq10$ for uncontrolled simulation (dots) and controlled simulation (crosses). }
\label{mean}
\end{center}
\end{figure}      
Fig.~\ref{mean} shows the uncontrolled and controlled mean energies for each controlled wavenumber pair. As predicted from Fig.~\ref{OLCL}, it is the lowest streamwise constant modes that are responsible for the largest contributions to the total perturbation energy of the flow. The controllers reduce mean modal energy for all modes controlled. The controller for wavenumber pair $\alpha=0,\beta=2$ reduced mean modal energy by over $50\%$.  
\begin{figure}
\begin{center}
\includegraphics[width=0.3\textwidth]{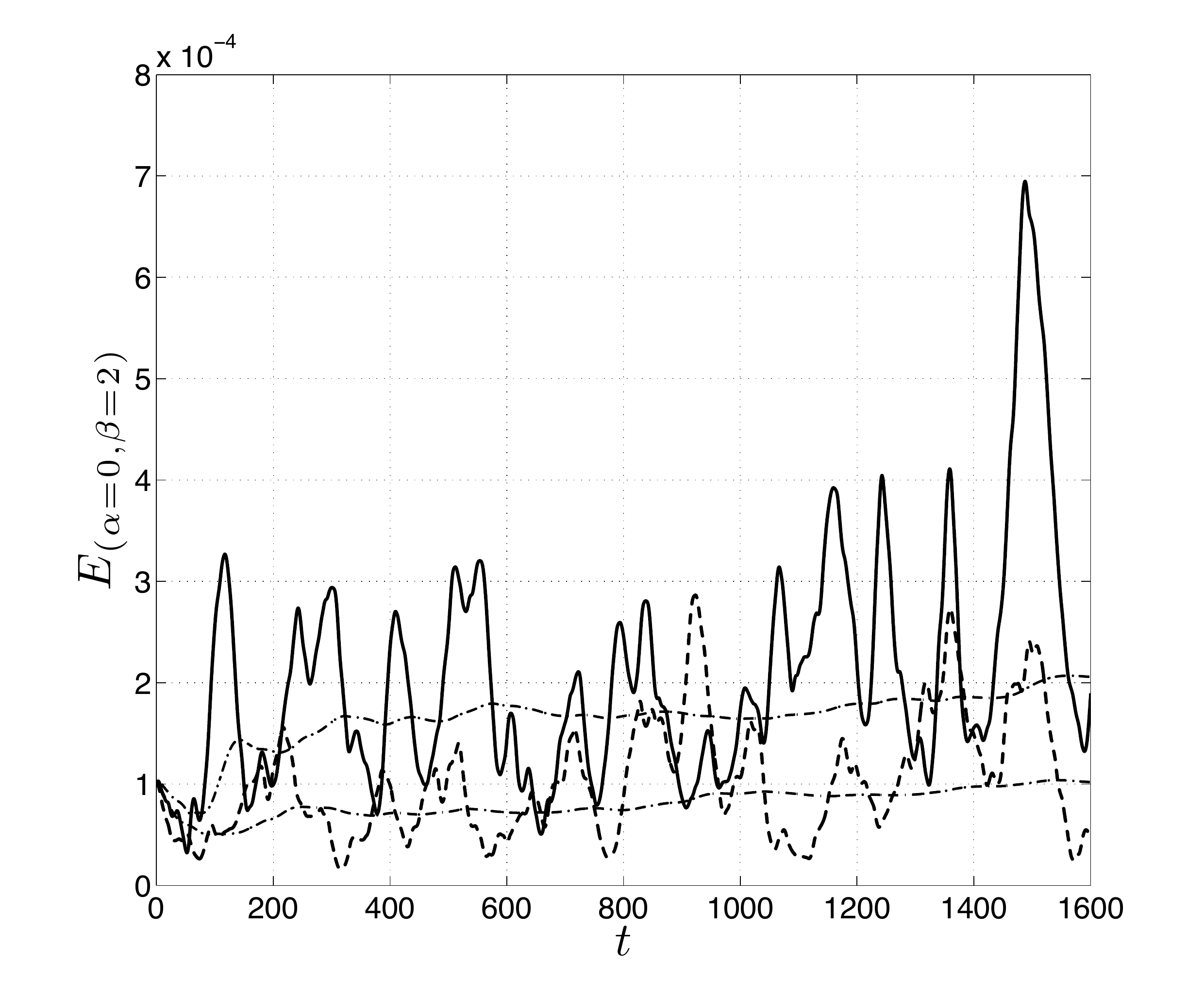}
\caption{Plots of modal energy for Fourier mode $\alpha=0,\beta=2$ from uncontrolled (thick line) and controlled (dashed line) simulations. Also included are dash-dot lines representing moving mean values.}
\label{en02}
\end{center}
\end{figure}      
Fig.~\ref{en02} shows plots of the energy time histories for mode $\alpha=0,\beta=2$. The effect of the controller is to reduce the size of the temporal peaks in energy for that mode, reducing the mean energy significantly. This illustrates how reducing $|\varepsilon|$ decreases the effect disturbances have on the energy production of a controlled mode. \\
\\
The total skin-friction drag at each wall is defined as:
\begin{equation}
\bar{D}_{\pm1} = \frac{1}{Re} \left | \left . \frac{\partial \langle{\mathrm{U}(\mathrm{y})}\rangle}{\partial \mathrm{y}}\right |_{\mathrm{y}=\pm1} \right |,
\end{equation}
where $\langle\mathrm{U}\rangle$ is the $\mathrm{x-z}$ spatial-mean streamwise velocity profile. Fig.~\ref{drag} shows time histories of total drag at both walls. 
The controllers achieved an overall reduction in temporal-mean total drag of $7\%$. There is some overlap of the uncontrolled and controlled drag time histories. However, the effect of the controllers is to reduce the temporal peaks in drag, bringing the mean values down. There is an initial large transient reduction in drag at the lower wall which corresponds to the large reduction in total perturbation energy seen in Fig.~\ref{energy}. Like the total perturbation energy, the drag at the lower wall recovers, but at a lower steady-state mean value. The transient responses at both walls appear very different. However, note that for both the uncontrolled and controlled flows, the temporal-mean drag values at both walls are very similar.   \\
\\
Comparing the power saved via drag reduction to the power spent on the control actuation provides insight into the efficiency of the controller. The mean power required to push a fluid along the streamwise direction of a channel is calculated from~\cite{Ricco13}: 
\begin{equation}
\mathcal{P}_\mathrm{x} :=  \mathrm{VU_b \overline{\left\langle\frac{\partial P}{\partial x} \right \rangle }},
\end{equation}
where $\mathrm{V}$ is the volume of the domain, $\mathrm{U_b}$ is bulk velocity, and $\mathrm{ \overline{\left\langle\frac{\partial P}{\partial x} \right \rangle }}$ is the temporal and spatial mean streamwise pressure gradient. The mean power of the control actuation is calculated from~\cite{BewleyMoin01}:
\begin{equation}
\mathcal{P}_\phi := \underbrace{\int_{\partial \Omega} \frac{1}{2} |\mathrm{v}^3| \ d \chi}_{\displaystyle \mathcal{P}_{\mathrm{v}^3}}  + \underbrace{ \int_{\partial \Omega}  |\mathrm{vp}| \  d \chi}_{\displaystyle \mathcal{P}_{\mathrm{vp}}}.
\label{ConPower}
\end{equation}
The first term on the right-hand-side of \eqref{ConPower} corresponds to the rate of addition of kinetic energy to the flow, and the second term corresponds to the rate of work done on the static pressure by the actuation. The control power efficiency is defined as:
\begin{equation}
\mathcal{P}_\% := 100 \left ( \frac{\mathcal{P}_\phi}{\Delta \mathcal{P}_\mathrm{x}} \right ),
\end{equation}
where $\Delta  \mathcal{P}_\mathrm{x} : = |\mathcal{P}_{\mathrm{x},\text{UNCON}}| - |\mathcal{P}_{\mathrm{x},\text{CON}}|$ is the power saved due to drag reduction. 
\begin{figure}
\centering
\begin{tabular}{c|c|c|c|c}
$\Delta \mathcal{P}_{\mathrm{x}}$ & $\mathcal{P}_{\mathrm{v}^3}$ & $\mathcal{P}_{\mathrm{vp}}$ & $\mathcal{P}_\phi$ & $\mathcal{P}_\%$ \\
\hline
\small{$1.6 \times$$10^{-2}$} & \small{$8.5 \times$$ 10^{-7}$} & \small{$3.9 \times$$ 10^{-4}$} & \small{$3.9 \times$$ 10^{-4}$} & \small{2.4}
\end{tabular}
\caption{Table summarising power saved, power spent, and the power efficiency of the controller, for $|\mathcal{P}_{\mathrm{x},\text{UNCON}}|= 2.2\times$$10^{-1}$. }
\label{PowerTab}
\end{figure}
The table in Fig.~\ref{PowerTab} summarises the mean power saved, the total mean power of the actuation, and the power efficiency. The controller has proven to be highly efficient with the power spent on the control being just 2\% of the power saved due to drag reduction.     \\
\\
Insight into how the controllers reduce energy and drag in the flow can be gained from visualisation of the flow fields. Fig.~\ref{swvelvort} shows colour maps of streamwise velocity overlaid with contour lines of streamwise vorticity on the $\mathrm{y}-\mathrm{z}$ plane for wavenumber pair $\alpha=0,\beta=2$, for the uncontrolled and controlled simulations.
The controller has reduced the perturbation energy for this mode by firstly, weakening the streamwise velocity perturbations near the walls. Secondly, the controller reduces the strength of streamwise vorticity close to the wall (presumably through counterrotation). Both the velocity perturbations and vortices are pushed away from the wall, as can be clearly seen on the top wall of the controlled flow.

\section{Conclusions}
The ten lowest streamwise constant Fourier modes of the linearised dynamics of channel flow were found to be responsible for the vast majority of system energy production; a result that coincides with previous research. Using a method based on that outlined by Sun et al.~\cite{Sun94}, controllers were synthesised to minimise the closed-loop energy bound $|\varepsilon_{L,K}|$ for these modes. Using high fidelity nonlinear simulations of turbulent channel flow, it was found that these controllers significantly reduced the mean perturbation energy of the modes they controlled. Their combined effort culminated in notable reductions in the mean total perturbation energy and mean total skin-friction drag of the flow. The new results presented in this paper demonstrate that when sensing and actuation is located at the walls, LTI control is only capable of \emph{restricting} the energy produced by each spatial mode. Sharma et al.~\cite{Sharma11} found that by sensing and actuating throughout the flow they could prevent energy production in all modes controlled. Controllers with alternative wall sensing and actuation to that used in the current work may enforce larger reductions in system energy bounds, this could result in even lower skin-friction drag.           \\
\\
In future work, this control method will be applied to supercritical Reynolds number channel flows. The linear dynamics of these flows are open-loop unstable to infinitesimal perturbations. Optimal sensing/actuation arrangements will also be investigated. Eventually, it is hoped to numerically implement passivity-based control on wall-bounded flow systems with more complex geometries using matrix-free methods~\cite{Theofilis11}. 
\label{Conclusions}

\begin{ack}                               The authors acknowledge support of a UK Engineering and Physical Sciences Research Council DTA.
\end{ack}

\bibliographystyle{plain}     
\bibliography{AutoBib}

\end{document}